# Uncertainty quantification for Multiphase-CFD simulations of bubbly flows: a machine learning-based Bayesian approach supported by high-resolution experiments

Yang Liu[1], Dewei Wang[2], Xiaodong Sun[3], Yang Liu[2], Nam Dinh[4], and Rui Hu[1]

[1]: Nuclear Science and Engineering Division, Argonne National Laboratory, United States
[2]: Department of Mechanical Engineering, Virginia Tech, United States
[3]: Department of Nuclear Engineering and Radiological Sciences, University of Michigan, United States
[4]: Department of Nuclear Engineering, North Carolina State University, United States

**Abstract**

In this paper, we develop a machine learning-based Bayesian approach to inversely quantify and reduce the uncertainties of the two-fluid model-based multiphase computational fluid dynamics (MCFD) for bubbly flow simulations. The proposed approach is supported by high-resolution two-phase flow measurement techniques, including double-sensor conductivity probes, high-speed imaging, and particle image velocimetry. Local distribution of key physical quantities of interest (QoIs), including void fraction and phasic velocities, are obtained to support the modular Bayesian inference. In the process, the epistemic uncertainties of the closure relations are inversely quantified, the aleatory uncertainties from stochastic fluctuation of the system are evaluated based on experimental uncertainty analysis. The combined uncertainties are then propagated through the MCFD solver to obtain uncertainties of QoIs, based on which probability-boxes are constructed for validation. The proposed approach relies on three machine learning methods: feedforward neural networks and principal component analysis for surrogate modeling, and Gaussian processes for model form uncertainty modeling. The whole process is implemented within the framework of open-source deep learning library PyTorch with graphics processing unit (GPU) acceleration, thus ensuring the efficiency of the computation. The results demonstrate that with the support of high-resolution data, the uncertainty of MCFD simulations can be significantly reduced.

1. Introduction

Two-phase flow and boiling heat transfer have a broad range of engineering applications. One of its most important application is for large-scale power systems, from coal and gas fired power stations to nuclear power plants. In nuclear engineering, the boiling crisis is critical to nuclear safety, while the void fraction has a significant influence on the reactivity of the reactor system. For nuclear power plants, accurately model the two-phase flow phenomena is of vital importance for both efficiency and safety. As risk-informed safety analysis start to be adopted by the regulatory authority, uncertainty and risk analysis become increasingly popular in the nuclear engineering community [1-4], making a comprehensive uncertainty quantification (UQ) of two-phase flow modeling a highly desirable task.

In current engineering practices, modeling two-phase flow with high-resolution for local flow details remains a difficult task. The main challenge is the treatment of the interface between two phases. Directly resolving the interfaces with methods like level set [5] require formidable computational resources. While considering the two-phase as a mixture such as the homogeneous model could oversimplify the complex interfacial exchange between the two phases.

One of the promising tools for high-resolution modeling of two-phase flow is the two-fluid model within the computational fluid dynamics (CFD) framework, i.e. multiphase-CFD (MCFD) [6, 7]. With MCFD, the interface between the liquid phase and gas phase are not explicitly resolved. Instead, the interfacial information is averaged, and a set of constitutive relations are introduced to make the averaged conservation equations solvable.

The MCFD solver relies on solving three ensemble- or time-averaged conservation equations for mass, momentum and energy [8]. The k-phasic mass conservation equation can be written as:

$$\frac{\partial(\alpha_k \rho_k)}{\partial t} + \nabla \cdot (\alpha_k \rho_k \mathbf{U}_k) = \Gamma_{ki} - \Gamma_{ik}, \tag{1}$$

where the two terms on the right-hand side represent the rate of mass exchanges between phases due to condensation and evaporation, which are modeled by constitutive relations.

The k-phasic momentum equation is given by

$$\frac{\partial(\alpha_k \rho_k \mathbf{U}_k)}{\partial t} + \nabla \cdot (\alpha_k \rho_k \mathbf{U}_k \mathbf{U}_k) = -\alpha_k \nabla p_k + \nabla \cdot [\alpha_k (\tau_k + \tau_k^t)] + \alpha_k \rho_k \mathbf{g} + \Gamma_{ki} \mathbf{U}_i - \Gamma_{ki} \mathbf{U}_k + \mathbf{M}_{ki}, \tag{2}$$

where $\mathbf{M}_{ki}$ represents the term of averaged interfacial momentum exchange, which are modeled by a set of interfacial force constitutive relations.

The k-phasic energy conservation equation in terms of the specific enthalpy can be given as:

$$\frac{\partial(\alpha_k \rho_k h_k)}{\partial t} + \nabla \cdot (\alpha_k \rho_k h_k \mathbf{U}_k) = \nabla \cdot \left[\alpha_k \left(\lambda_k \nabla T_k - \frac{\mu_k}{\Pr_k^t} \nabla h_k\right)\right] + \alpha_k \frac{Dp}{Dt} + \Gamma_{ki} h_i - \Gamma_{ik} h_k + q_k, \tag{3}$$

where the wall boiling heat transfer $q_k$ is modeled by a set of constitutive relations.

Most of the constitutive relations MCFD solver relying on are empirical or semi-empirical correlations, which usually evolve multiple empirical parameters that need to be set up for the simulation. A detailed characterization of the MCFD constitutive relations and the role of empirical parameters can be found in [9]. These parameters constitute a major source of uncertainty for the MCFD simulations that need to be quantified so that the simulation results can be used for comprehensive risk analysis. This UQ process is essentially an inverse problem and relies on experimental measurements to support the inference process.

The legacy two-phase datasets have several drawbacks that makes it not suitable to support the UQ of high-resolution MCFD simulation. First, many legacy data do not contain all the necessary flow features, i.e. quantities of interest (QoIs), as the legacy experiments either focus on gas phase flow or liquid phase flow, so the legacy data with void fraction measurement may not have the corresponding liquid velocity measurements available. Second, the legacy two-phase flow data were usually low-resolution that are not compatible to the high-resolution nature of the MCFD simulations. Third, many legacy experiments did not have clear information on the inlet and boundary conditions, which makes it difficult to setup simulations that accurately reflect the experimental conditions. Last, the legacy experiment usually did not report measurement uncertainty, or the uncertainties were not adequately quantified. In this sense, we need two-phase flow experimental datasets that can overcome all these shortcomings for the UQ of the MCFD simulations.

In the past decade, advanced techniques have been developed for two-phase flow measurement for both gas and liquid phases. These techniques have demonstrated successful applications for bubbly flow under various conditions. For gas-phase measurement, a high-speed imaging system employing multiple high-speed cameras and the associated image processing algorithms have been established [10] and demonstrated good performance. For liquid-phase measurement, the integration of Particle Image Velocimetry (PIV) and Planar Laser-Induced Fluorescence (PLIF) techniques [11, 12], associated with the methods to minimize the effect of light distortion on bubble surfaces have been developed to measure liquid-phase velocity field in bubbly flow. These two techniques, along with the conductivity probe, a mature and robust technique for measuring local gas phase QoIs [13], could be integrated to develop a comprehensive database for the UQ of MCFD simulations.

Quantifying the uncertainty of MCFD simulation with experimental data is an inverse problem, i.e. we rely on the measurements of the QoIs to quantify the uncertainty of the factors that predict them. Compared to forward UQ problem where uncertainties are directly propagated through the computational model to generate uncertainty evaluation [14, 15], inverse UQ is considered to be more challenging and requires more sophisticated treatments[16-19]. A widely adopted approach for inverse problems of complicated engineering systems is the Bayesian inference, which usually requires tens of thousands of evaluations of the computational model to generate posterior uncertainty distribution of the parameters under investigation. In this sense, a surrogate model of the original model is usually required for the inference. Bayesian inverse UQ based on surrogate modeling is first introduced in [20] and has been tested in many complex engineering problems with successful applications [17, 21-24]. In these applications, surrogate models are developed using the Gaussian processes (GPs), a machine learning method that has applications on a variety of topics [25-27]. For problems with high-dimensional outputs, dimensionality reduction is usually required so that the surrogate model can be constructed on a reduced size of outputs [28, 29].

In this work, we propose a modular Bayesian inference approach for the inverse UQ of MCFD simulation. The Bayesian inference is supported by three widely used machine learning methods: we employ feedforward neural network (FNN) and principal component analysis (PCA) to construct a surrogate model for the MCFD solver; we rely on Gaussian processes (GPs) to evaluate the model form uncertainty. We leverage a comprehensive two-phase bubbly flow experimental database obtained with multiple experimental techniques to support this UQ work.

There are two major novelties in terms of the methodology of the proposed approach. Firstly, we integrate the dimensionality reduction (with PCA) into the surrogate model (in the form of FNN) training process. Compared to a similar pioneering work [28], where dimensionality reduction and surrogate model are developed separately, this integrated process is able to reduce both the surrogate model error and the additional error introduced by the dimensionality reduction process. Secondly, in this work, we not only quantify and reduce the epistemic uncertainty of the MCFD model, but also evaluate the irreducible aleatoric uncertainty of the stochastic fluctuation of the bubbly flow system. Based on the combination of epistemic uncertainty and aleatoric uncertainty, we are able to construct probability-boxes (p-boxes) that can be used for different purposes in the context of uncertainty and risk analysis. Furthermore, the whole approach is based on open-source deep learning library PyTorch [30] with the support of graphical processor unit (GPU), ensuring a streamlined implementation with outstanding computational efficiency.

The remaining of this paper is organized as follows: Section 2 introduces the proposed machine learning-based Bayesian approach; Section 3 provides details on the high-resolution experiments and the associated uncertainty analysis; Section 4 defines the problem of bubbly flow investigated in this work; Section 5 discusses the results; and Section 6 finally delivers summary remarks of our work.

## 2. Machine learning-based modular Bayesian inference

For a general computational model that dependent on spatial location $x$, inlet condition $\eta$, and constitutive closure parameter $\theta$, the relationship between the model predictions $\mathbf{y}^M(x, \eta, \theta)$ and experimental measurements $\mathbf{y}^E(x, \eta)$ can be expressed as

$$\mathbf{y}^E(x, \eta) = \mathbf{y}^M(x, \eta, \theta) + \delta(x) + \varepsilon(x), \tag{4}$$

where $\delta(x)$ is the model form uncertainty, and $\varepsilon(x)$ is the measurement uncertainty. Within the Bayesian framework, $\theta$ is also considered as random variables with uncertainty distribution. In this work, we developed a machine learning-based Bayesian approach to inversely quantify the uncertainty of MCFD simulations, the proposed approach relies on experimental measurements to do inference and considers the three sources of uncertainties comprehensively.

### 2.1 Surrogate modeling

Running the original MCFD simulation a few thousand times for the Bayesian inference would be formidably computationally expensive. In this sense, a surrogate model that could serve as an accurate representation of the original MCFD simulation but with computational efficiency is desired. Considering the high-dimensionality of the MCFD outputs, dimensionality reduction is needed before the surrogate model is constructed, so that the overall computational cost can be reduced.

In this work, we utilize the Principal Component Analysis (PCA) method to reduce the dimensionality of the MCFD outputs. We then employ the Feedforward Neural Network (FNN) to construct the surrogate model based on the reduced MCFD outputs subspace. A training strategy is developed that combines PCA with FNN so that both the FNN error and PCA-induced error can be minimized. The whole surrogate modeling process is based on *M* original MCFD simulations with perturbed constitutive relation parameters generated from their prior distributions. In practice, we employ the Singular Value Decomposition (SVD), an algorithm with superior numerical stability, to identify the principal components.

We consider three QoIs in this work, i.e. void fraction, gas phase velocity, and liquid phase velocity. The experimental measurements of the three QoIs are performed on a horizontal plane of a vertical rectangular channel, as will be discussed in detail in Section 3. The results of the three QoIs on the measurement plane from each MCFD simulation is extracted, reshaped, and concatenated to form a column vector $\boldsymbol{q}$. A total $M$ column vectors are combined to form a matrix $\boldsymbol{Q}$. SVD is performed on $\boldsymbol{Q}$ and decompose it into three matrices:

$$\boldsymbol{Q} = \mathbf{U}\Sigma\mathbf{V}^T, \tag{5}$$

where $\Sigma$ is a diagonal matrix whose entry values corresponding to the root of the eigenvalues of $\boldsymbol{QQ}^T$; $\mathbf{U}$ is a unitary matrix, whose column vectors are termed the Principal Components (PCs), with the equal size of the column vector of $\boldsymbol{Q}$, and $\boldsymbol{V}$ is a $M \times M$ square unitary matrix. We retain the first $k$ PCs from $\mathbf{U}$ to form a matrix $\boldsymbol{\Phi}$. $\boldsymbol{\Phi}$ is chosen in a way that it can account for at least 90% of the total variance of $\boldsymbol{Q}$. $\boldsymbol{\Phi}$ is then used to map the column vector $\boldsymbol{q}$ to the PC subspace $\boldsymbol{y}$, so its dimensionality is significantly reduced (from a few thousand mesh grids to 10-30 PCs). Moreover, results on PC subspace can be recovered to its original physical space through $\boldsymbol{\Phi}$. Such a dimensionality reduction process is illustrated in Figure 1.

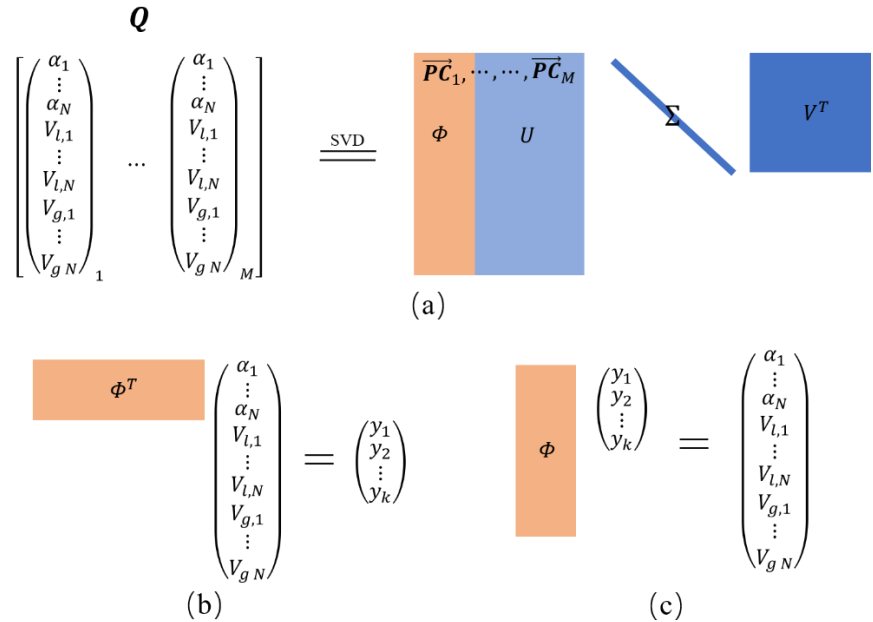

**Figure 1. (a). Obtaining principal components through Singular Value Decomposition; (b) Mapping MCFD simulation results from physical space to principal component subspace; (c) Recover results on principal component subspace to original physical space**

Based on the PC subspace, we further use the FNN to construct a surrogate model. The FNN consists of an input layer, multiple hidden layers, and an output layer. It can be regarded as a process where the input features go through a series of nonlinear transformations, which is controlled by learnable parameters weights $\boldsymbol{W}$ and biases $\boldsymbol{b}$, to predict its outputs. Such a prediction process needs to be properly trained with given input-output data pairs before it can make reasonable predictions. To train an FNN, a loss function $L(\hat{\boldsymbol{y}}, \boldsymbol{y})$ needs to be defined to measure the discrepancy between its prediction $\hat{y}$ and the real data $y$. The Root-of-Mean-Square based on $L_2$ norm is widely used for regression related problems:

$$L(\hat{y}, y) = \sqrt{\|\hat{y} - y\|_2} = \sqrt{\frac{1}{m}\sum_{i=1}^{m}(\hat{y}_i - y_i)^2} \tag{6}$$

With the loss function defined, the gradients of FNN prediction error with regard to the weights and biases of each hidden layer can be computed through the backpropagation algorithm [31]. Through this way, these weights and biases can be graduated updated to reach a global optimum where FNN prediction has a minimized error for the whole input-output data pairs. More detailed discussions of implementing FNN to flow and heat transfer problems can be found in [32-34].

In this work, the FNN takes the constitutive relation parameters $\theta$ as inputs to predict the MCFD results on PC subspace $y$, which is then mapped back to the physical space. The loss function is constructed based on the results of the original physical space, as illustrated in Figure 2. In this way, both errors, including the error of FNN prediction on PC subspace $y$, and the error of mapping PC subspace $y$ to original physical space, are considered in the training process. The detailed training process, including efforts to mitigate the overfitting issue, will be discussed in Section 5.

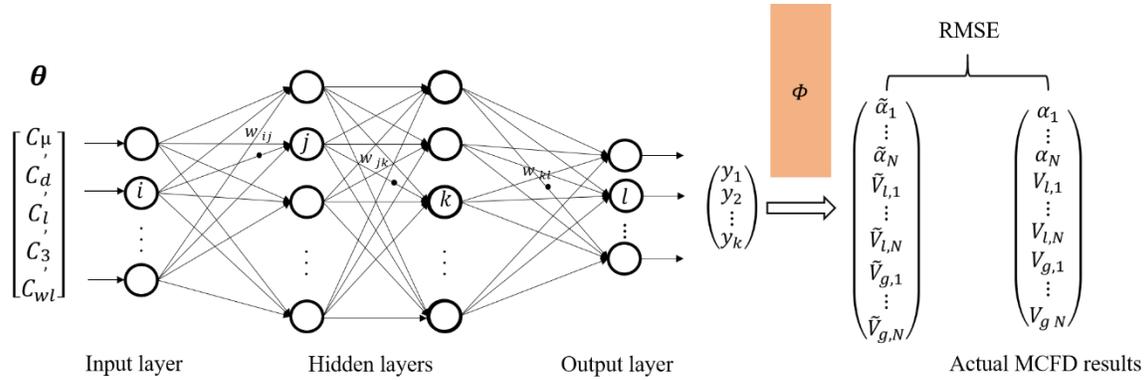

**Figure 2. Surrogate modeling with a combination of PCA and FNN**

**2.2 Modular Bayesian inference**

After the surrogate model for MCFD simulation is constructed, the model form uncertainty term $\delta(x)$ that is independent of parameter $\theta$ still needs to be modeled for a comprehensive UQ of MCFD predictions. In this paper, $\delta(x)$ is modeled as Gaussian Processes (GPs) $\mathcal{GP}(0, C)$. In this work, the covariance function $C(x, x')$ is constructed using the radial basis function kernel:

$$C(\pmb{x}, \pmb{x}') = \sigma^2 \exp\left(-\frac{|\pmb{x} - \pmb{x}'|^2}{2\ell^2}\right), \tag{7}$$

in here $\sigma^2$ and $\ell$ are hyperparameters of the GP that represent the variance and length scale respectively. The detailed discussion of using GP for regression problem can be found in [35]. Choosing GP to represent the prior of an unknown function $\delta(\pmb{x})$ has justification that can be found in [20]. In this work, $\delta(\pmb{x})$ is trained by the discrepancy between MCFD model predictions and experiment data at relatively sparse measurement sites. Once $\mathcal{GP}(0, C)$ is trained, it can be used to predict at any arbitrary location, so the model form uncertainty at every grid point of the MCFD prediction can be evaluated. A detailed implementation of GP for modeling $\delta(\pmb{x})$ on a similar problem can be found in [36].

With surrogate model and model form uncertainty term developed, the inverse UQ can be performed within the Bayesian framework:

$$p_{post}(\pmb{\theta}|\pmb{y}^E, \pmb{y}^M) = \frac{L(\pmb{y}^E, \pmb{y}^M|\pmb{\theta})p_{prior}(\pmb{\theta})}{p(\pmb{y}^E, \pmb{y}^M)} = \frac{L(\pmb{y}^E, \pmb{y}^M|\pmb{\theta})p_{prior}(\pmb{\theta})}{\int_{\mathbb{R}^p} p(\pmb{y}^E, \pmb{y}^M|\pmb{\theta})p_{prior}(\pmb{\theta})d\pmb{\theta}}, \tag{8}$$

where $p_{post}(\pmb{\theta}|\pmb{y}^E, \pmb{y}^M)$ is the posterior uncertainty of $\pmb{\theta}$ given the experimental measurements $\pmb{y}^E$ and model predictions $\pmb{y}^M$, $L(\pmb{y}^E, \pmb{y}^M|\pmb{\theta})$ is the likelihood function, and $p_{prior}(\pmb{\theta})$ is the prior uncertainty of $\pmb{\theta}$. $\mathbb{R}^p$ means the model parameter space has a dimensionality of $p$.

The likelihood function measures the "likelihood" of observing the experimental measurements $\pmb{y}^E$ and model prediction $\pmb{y}^M$, given specific values of $\pmb{\theta}$, and can be expressed as:

$$L(\pmb{y}^E, \pmb{y}^M|\pmb{\theta}) \propto \exp\left(-\frac{1}{2}[\pmb{y}^E - \pmb{y}^M - \delta]^T \Sigma^{-1} [\pmb{y}^E - \pmb{y}^M - \delta]\right), \tag{9}$$

where $\Sigma$ is a matrix that combines the covariance matrices of $\pmb{y}^M$, $\delta$, and $\varepsilon$.

For most engineering problems the integral over the space $\mathbb{R}^p$ is intractable, so $p_{post}(\pmb{\theta}|\pmb{y}^E, \pmb{y}^M)$ cannot be computed directly. A widely used approach to obtain the posterior density without explicitly do the integration over the whole space of $\pmb{\theta}$ is the Markov chain Monte Carlo (MCMC) sampling, a method that constructs Markov chains whose stationary distribution is the posterior density of $\pmb{\theta}$.

Technically speaking, one could lump all the relevant information into Bayesian inference to infer all the parameters through one process. The parameters include model parameter $\pmb{\theta}$, and parameters/hyperparameters of FNN and GPs. Such an approach is termed as full Bayesian approach [37]. Full Bayesian is theoretically straightforward and arguably superior, but could lead to serious identifiability issue as the parameter space that the Bayesian inference deal with would be very high dimensional. Especially in this case where FNN could involve thousands of parameters in the form of weights and biases. A more practical approach, especially for complex engineering problems, is the modular Bayesian approach [17]. Such an approach trains the surrogate model and model form uncertainty in two separate modules, then plug in the two models as known terms in the Bayesian inference that only deal with the model parameter $\boldsymbol{\theta}$. Modular Bayesian demonstrate great potential for complex engineering problems and has successful applications on two-phase flow and boiling heat transfer in one of our previous works [36]. Based on this reason, we adopted the modular Bayesian approach in this work, the specific steps for conducting the inverse UQ through this approach is illustrated in Figure 3.

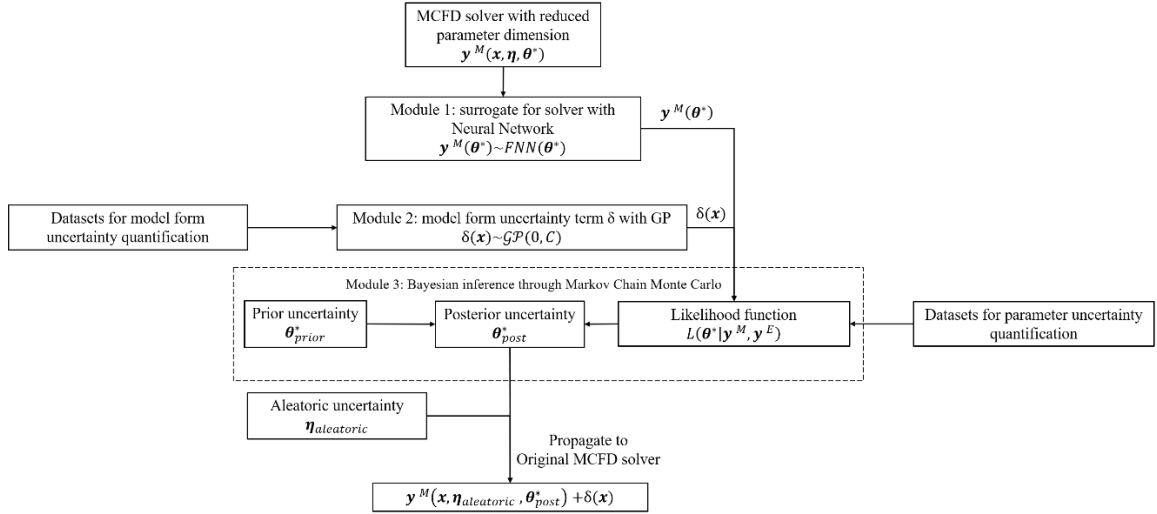

**Figure 3. Modular Bayesian for inverse UQ of MCFD simulations**

## 2.3 Results interpretation: HPD intervals and P-boxes

MCMC will generate posterior samples of model parameter $\boldsymbol{\theta}$, based on which the posterior distribution $p_{post}(\boldsymbol{\theta}|\mathbf{y}^E, \mathbf{y}^M)$ can be constructed. On the one hand, we need the full multi-dimensional posteriors to study the correlation between different parameters. On the other hand, we need a more concise summary of the posterior of each parameter in the form of marginal credible intervals to better evaluate their individual impact on the solver prediction. Based on the posterior samples, the marginal credible interval is evaluated with the Highest Posterior Density (HPD) method. HPD interval has two good properties that making it suitable for the credible interval evaluation, i.e. (1) any sample within the interval has a higher density than any other point outside; and (2) for any credible interval with given probability (1-α)×100%, HPD interval is always the shortest.

Based on the obtained posterior distributions $p_{post}(\mathbf{\theta}|\mathbf{y}^E, \mathbf{y}^M)$, the uncertainty of MCFD predictions can be quantified. In this work, we use the obtained MCFD prediction uncertainties to construct probability-box (p-boxes) [38]. P-box is an approach to describe imprecise probability, which expresses both epistemic uncertainty (i.e. uncertainty due to lack of knowledge) and aleatoric uncertainty (i.e. uncertainty due to the stochastic nature of the physical process that is irreducible) in a way that does not confound the two. In this work, epistemic uncertainties include both model form uncertainty and model parameter uncertainty, while aleatoric uncertainty consists of the fluctuation of the inlet conditions. In a p-box, the horizontal range stands for the prediction uncertainty due to the epistemic uncertainty while the cumulative probability stands for the aleatoric uncertainty. The constructed p-box can then be used for different purposes, including risk assessment [39], reliability analysis [40], and sensitivity analysis [41].

## 3. Data support from high-resolution experiments

### 3.1 Experimental facility

In this work, we rely on the experimental facility at Virginia Tech to develop a comprehensive two-phase flow database with high-resolution measurements to support the inverse UQ work.

The schematic of the experimental test loop is depicted in Figure 4. This facility is designed for air-water two-phase flow at room temperature and near atmospheric pressure. The test section is 3 m tall vertical channel with a rectangular cross section of 30 mm × 10 mm. The hydraulic diameter of this test section is $D_h$ = 15 mm. Three instrumentation ports are located at normalized axial distances of $z/D_h$ = 8.8, 72.4, and 136 from the two-phase injector, where measurements are performed. The gas is injected on the side at the test section bottom through a pair of opposite facing aluminum plate, each has five 200 μm holes. Three experimental techniques are utilized to measure key QoIs for a two-phase bubbly flow system with local detail and high-resolution, including two phasic velocities and void fraction. Besides that, the bubble size measured at Port 2 is also used as an inlet flow condition.

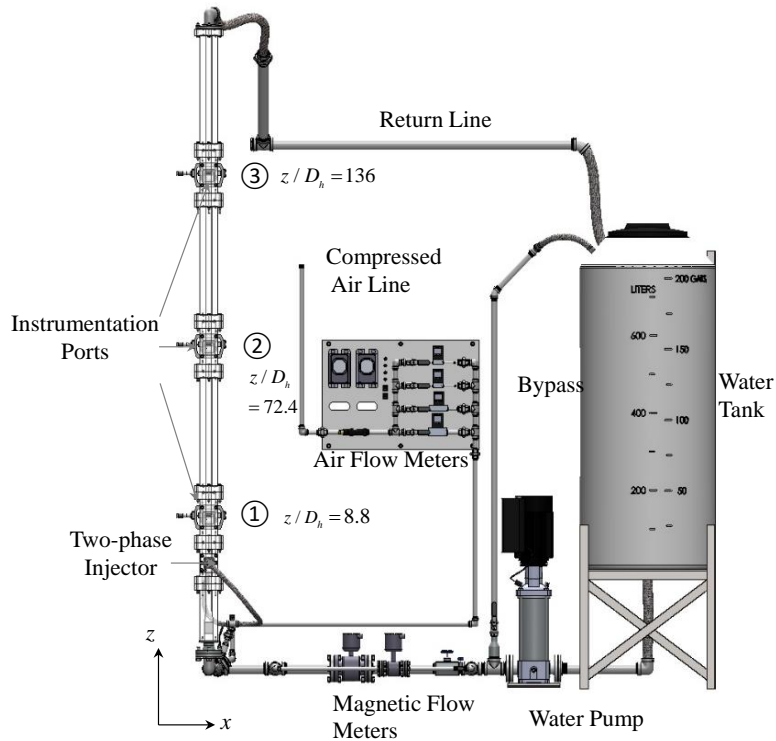

**Figure 4. Schematic of rectangular pipe test loop**

### 3.2 Measurement techniques

*Conductivity probe*

The conductivity probe is used to obtain local time-averaged two-phase flow parameters. The working principle is shown in Figure 5 (a). As a bubble passing by the sensor tips, it would be registered as a voltage changes, ideally a square wave. However, actual signals are different from the ideal ones due to issues like white noise, electromagnetic interference, cross-talk (or ghosting), and finite response time of the electronics, etc. A signal processing program is developed to process the raw voltage signals to extract the time information needed to calculate time-averaged two-phase flow parameters such as void fraction, velocity, and superficial gas velocity [13, 42].

In the actual measurements, the double sensor probe is placed on a two-way linear stage to move to a designated location in the cross-sectional plane of the test section. Due to the finite size of the probe casing, the measurement points vary from 0 to 14 mm along the x (length) direction, and from 0 to -3.9 mm along the y (width) direction. After the probe signal is recorded at one location, the probe is traversed to a neighboring point for data acquisition until all the black dots shown in Figure 5 (b) are covered. For each test run, probe measurements are performed at $8 \times 4$ locations for all three instrumentation ports. With a symmetric assumption, the distributions of time-averaged parameters at symmetric locations (blank dots) are mirrored from the measurements over a quarter of the cross section of the rectangular channel. After signals are processed, the local time-averaged parameters including void fraction and gas-phase velocity can be obtained.

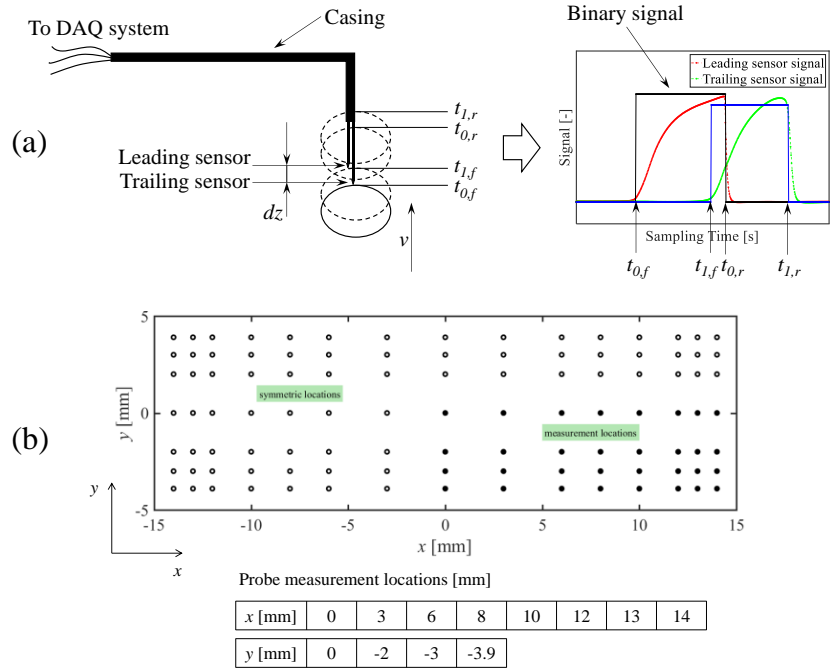

**Figure 5. Gas phase measurement with conductivity probe (a). Working principle of a double-sensor conductivity probe; (b). Measurement locations**

*PIV-PLIF system*

The PIV system used in this study is combined with the planar laser-induced fluorescence (PLIF) technique, an optical phase separation method, as depicted in Figure 6. Fluorescent particles and optical filtration are applied to minimize the effect of bubble surfaces and thus improve the image quality [11, 12]. With the PIV-PLIF system, the bubble-induced light distortion in the PIV measurement can be mitigated. Additional uncertainty analysis through ray optics modeling also confirms the distortion from bubble on PIV measurement can be filtered out through multi-pass post-processing technique [43], thus PIV-PLIF can be regarded as a valid tool for liquid phase measurement in two-phase flow systems.

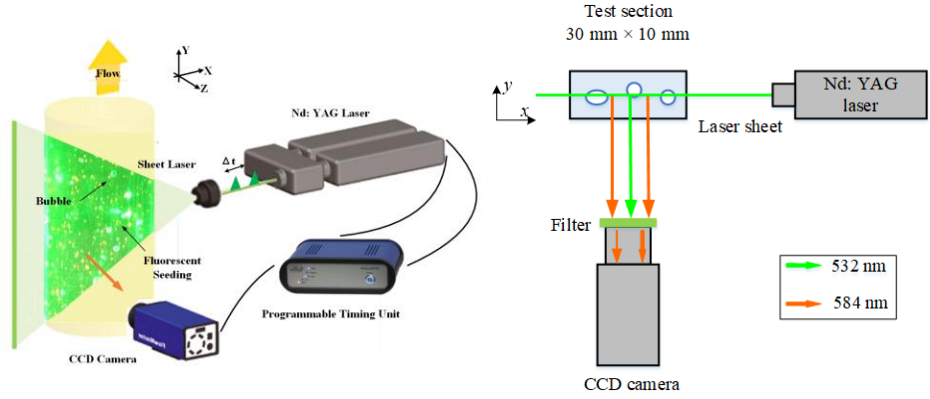

**Figure 6. PIV-PLIF system for bubbly flow measurement**

*High-speed camera system*

Highspeed camera is a mature tool to visualize two-phase flow interfacial structure. With image processing algorithms, the high-speed camera images can be automatically processed to extract bubbly flow QoIs with high temporal and spatial resolutions [10]. In this paper, the results obtained from high-speed camera are mainly used to validate the conductivity probe measurements.

**3.3 Experimental measurements and uncertainty analysis**

In this work, a dataset containing two different inlet conditions is measured with the above-mentioned techniques. The first case is measured at $j_f$=3.7 m/s, $j_g$=0.05 m/s (here $j$ denotes superficial velocity), which is denoted as Condition 1. The second case is measured $j_f$=1.0 m/s, $j_g$=0.2 m/s, which is denoted as Condition 2. Both cases are in the bubbly flow regime, so it is reasonable to assume they are governed by the same set of constitutive relations in the MCFD simulation.

As shown in Eq.(9), the measurement uncertainty is considered in the likelihood function in the Bayesian inference. Furthermore, the measurement uncertainty also influences the GPs that used to describe the model form uncertainty $\delta(x)$, as will be discussed in Section 5.2. In this work, we evaluate the measurement uncertainty through detailed analysis. For simplicity, in the remaining of this paper, we interpret a measurement uncertainty as the 95% confidence interval (CI) of that measurement. For example, when we say "the uncertainty of gas phase velocity measurement is 1%", we mean "the 95% CI of that measurement is bounded by 0.99-1.01 of the measured value."

As a major tool for gas-phase measurement, the uncertainty of the double sensor conductivity probe has been thoroughly investigated in a previous study [44]. For void fraction $\alpha$ measurement, its uncertainty is not influenced by either the probe configuration or bubble velocity fluctuation, rather the major uncertainty source is the interfacial time measurement error. With optimized signal processing algorithm, the uncertainty of local void fraction is around 5% for the dataset used in this study.

For gas phase velocity $V_g$ measurement, the conductivity probe measurement uncertainty is evaluated by comparing it with the line-averaged velocity measured by the high-speed imaging technique [10, 13]. By integrating over the cross-sectional plane, probe measurements can be compared to the high-speed imaging results, the percentage errors are summarized in Table 1. For Run 1, the uncertainty of $V_g$ measurement varies from 7.8% to 1.9% at different ports, and for Run 2, the uncertainty varies from 3.8% to 0.1%. The uncertainty analysis reflects a fact that gas velocity measurement has higher accuracy as bubbly flow is fully developed.

For liquid phase velocity $V_l$ measurement, the PIV-PLIF measurement uncertainty is converted to the uncertainty in the superficial liquid velocity. In this case, one can integrate the local PIV-PLIF measurements over the cross-sectional plane and compare the integrated value to the flow meter readings, where the magnetic flow meters should be very accurate as it measures the single-phase flow and has been well calibrated. Through the comparison, we found that the uncertainty for liquid velocity is 5% in Run 1. While for Run 2, the uncertainty is around 8%.

For Sauter mean diameter $D_s$, its uncertainty can be evaluated from computing the variation of all the bubbles measured by the conductivity probe. In this study, $D_s$ is measured and averaged for a single value in each Run, as $D_s$ demonstrates only a small variation in the test section. For Run 1, the averaged $D_s$ is 3.3 mm, with its uncertainty around 8%. For Run 2, the averaged $D_s$ is 4.9 mm, with its uncertainty around 10%. These obtained uncertainty information are summarized in Table 1.

Table 1. 95% CI of measurement uncertainties in terms of measurement value

|  |  | Port 1 | Port 2 | Port 3 |
|---|---|---|---|---|
| Uncertainty in $V_g$ [%] | Run 1 | 7.1 | 7.8 | 1.9 |
|  | Run 2 | 3.8 | 1.4 | 0.1 |
| Uncertainty in $V_l$ [%] | Run 1 | 1.7 | 1.5 | 4.3 |
|  | Run 2 | 7.2 | 8.0 | 8.4 |
| Uncertainty in $\alpha$ [%] | Run 1 | 5 | 5 | 5 |
|  | Run 2 | 5 | 5 | 5 |
| Uncertainty in $D_s$ [%] | Run 1 | 8 | 8 | 8 |
|  | Run 2 | 10 | 10 | 10 |

## 4. Problem setup

### 4.1 Numerical case setup

As the inverse UQ is based on surrogate modeling, the first step is to run simulations and extract QoIs for the surrogate model construction. The commercial CFD package STAR-CCM+ is used for MCFD simulations of the bubbly flow system. The computational domain is set to be consistent with the experimental test section, a 30mm×10mm vertical square channel. To have an accurate description of the input condition, the inlet of the simulation is set at the Instrumentation Port 2 in the experimental facility (z/D = 72.4). The phasic velocity, void fraction, and bubble size at the inlet can therefore be prescribed accurately with the measurements. The measurements at the Instrumentation Port 3 (z/D = 136) are used for the inverse UQ, so the outlet boundary of the simulation is set at 0.1 m above the Port 3 to minimize the influence of boundary on the Port 3 results. The total length of the channel in the simulation is 1.372 m. Hexahedral grids are constructed for the simulation domain and a cross-sectional view along with the experimental measurement sites is shown in Figure 7. It is worthy to note that there are 2240 meshes on the cross-section of the computation domain, which means each simulation has 2240 outputs for one single QoI. For three QoIs that involved in the UQ, there would be 6720 outputs, which is a high-dimensional problem. As discussed in Section 2.1, PCA is used to reduce the dimensionality, then combined with FNN to construct the surrogate model.

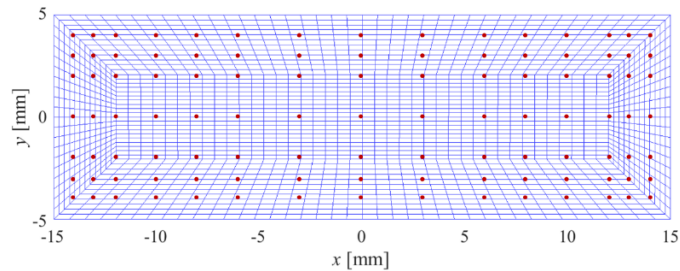

**Figure 7. Cross sectional view of mesh setup for MCFD simulation and the corresponding experimental measurement locations (marked by red dots)**

It should be noted that the mesh setup also constitutes a numerical uncertainty source in the MCFD simulation [45]. In this work, a mesh convergence study is performed, and we consider the mesh-induced numerical uncertainty is negligible compared to the uncertainty of constitutive relations.

### 4.2 Constitutive relations for inverse UQ

After the setup of mesh and inlet/boundary conditions for the computation domain, constitutive relation parameters can be sampled to run simulations. In this work, we rely on 500 simulations with perturbed parameters from constitutive relations to construct the surrogate model. It has been observed in multiple experiments that the interfacial forces and liquid turbulence have significant influences on the lateral distribution of the void fraction and phasic velocities [46]. In this paper, constitutive relations for the interfacial forces and liquid turbulence are considered in the inverse UQ process.

In a typical MCFD solver, the interfacial forces are described by multiple constitutive relations, i.e., drag, lift, turbulence dispersion, wall lubrication, and virtual mass forces. In this work, we do not consider the uncertainty of turbulence dispersion force and virtual mass force, as the former is strongly dependent on drag force, while the latter is usually one order magnitude smaller than other major interfacial forces. A brief summary of the interfacial forces investigated in this work is provided in

Table 2.

**Table 2. Constitutive relations for interfacial forces**

| Interfacial force | Expression |
| --- | --- |
| Drag force [8] | $\mathbf{M}_g^D = -\dfrac{3}{4}\dfrac{C_d}{D_s}\rho_l\alpha\|\mathbf{U}_g - \mathbf{U}_l\|(\mathbf{U}_g - \mathbf{U}_l)$ |
| Lift force [8] | $\mathbf{M}_g^L = C_l\rho_l\alpha(\mathbf{U}_g - \mathbf{U}_l) \times (\nabla \times \mathbf{U}_g)$ |
| Wall lubrication force [47] | $\mathbf{M}_g^{WL} = -f_{WL}(C_{wl}, y_w)\alpha\rho_l \dfrac{\|\mathbf{U}_r - (\mathbf{U}_r \cdot \mathbf{n}_w)\mathbf{n}_w\|^2}{D_s}\mathbf{n}_w$, $f_{WL}(C_{wl}, y_w) = \max\left(-0.2C_{wl} + (\dfrac{C_{wl}}{y_w})D_s, 0\right)$ |

The liquid turbulence is typically modeled using the turbulence models developed for single-phase turbulent flow with the consideration of bubble-induced turbulence. In this study, we use the $k - \varepsilon$ model as the turbulence model for the liquid phase, which solves two differential transport equations in order to determine the turbulent kinetic energy $k$ and the turbulent dissipation $\varepsilon$ for the liquid phase [48]:

$$\mu_k^T = \rho_k C_\mu \frac{k^2}{\varepsilon} \tag{10}$$

$$\frac{\partial(\alpha_k\rho_k k_k)}{\partial t} + \nabla \cdot (\alpha_k\rho_k \mathbf{U}_k k_k) = \nabla \cdot \left[\alpha_k\left(\mu_k + \frac{\mu_k^T}{\sigma_{k_k}}\right)\nabla k_k\right] + \alpha_k P - \alpha_k\rho_k\varepsilon_k + \alpha_k\Phi_k \tag{11}$$

$$\frac{\partial(\alpha_k\rho_k\varepsilon_k)}{\partial t} + \nabla \cdot (\alpha_k\rho_k \mathbf{U}_k \varepsilon_k) = \nabla \cdot \left[\alpha_k\left(\mu_k + \frac{\mu_k^T}{\sigma_{\varepsilon k}}\right)\nabla\varepsilon_k\right] + \alpha_k C_{\varepsilon 1}\frac{\varepsilon_k}{k_k}P - \alpha_k\rho_k C_{\varepsilon 2}\frac{\varepsilon_k}{k_k} + \alpha_k\Phi_{\varepsilon k} \tag{12}$$

In Eqs. (11) and (12), $P$ is the production of the shear-induced turbulent kinetic energy; $\Phi_k$ and $\Phi_{\varepsilon k}$ are the source terms due to the effects of the dispersed phase on the turbulence, and are modeled by the bubble-induced turbulence relation.

When considering the bubble-induced turbulence, $\Phi_k$ and $\Phi_{\varepsilon k}$ need to be calculated. In this study, the Troshko-Hassan model [49] is applied. This model assumes that all the work that is done by the drag force is an unconditionally positive production of continuous-phase pseudo-turbulence. It also assumes that this strong turbulence source is dissipated locally using a particle relaxation time scale. Thus the source term $\Phi_k$ can be expressed as:

$$\Phi_k = \alpha M^D |\boldsymbol{U}_r|, \tag{13}$$

where $M^D$ is the drag force and $\boldsymbol{U}_r$ is the relative velocity between the two phases. Based on $\Phi_k$, $\Phi_{\varepsilon k}$ can be calculated as

$$\Phi_{\varepsilon k} = \frac{C_3 \Phi_k}{t_b}, \tag{14}$$

where $C_3$ is another calibration coefficient with a default value 0.45 and $t_b$ is the bubble pseudo-turbulence dissipation relaxation time. The standard wall function widely used in turbulence modeling has been applied to the channel walls, with non-slip condition assumed for the liquid phase and a slip condition assumed for the gas phase.

The parameters from the aforementioned constitutive relations that are considered in the current inverse UQ process are summarized in Table 3. A nominal value is assigned for each parameter which would be used to generate a baseline model for model form uncertainty evaluation. These nominal values are either the default values in STAR-CCM+'s model setup or determined based on authors' previous experience on similar simulations. In the inverse UQ process, we conservatively assume "non-informative" priors for these parameters and make them follow uniform distributions.

**Table 3. Prior uncertainty setup for parameters in constitutive relations**

| Parameter | Constitutive relation | Distribution type | Distribution range | Nominal value |
|---|---|---|---|---|
| $C_\mu$ | Turbulence viscosity | Uniform | [0.07, 0.12] | 0.09 |
| $C_d$ | Drag force | Uniform | [0.44, 0.95] | 0.5 |
| $C_l$ | Lift force | Uniform | [-0.1, 0.1] | -0.01 |
| $C_3$ | Bubble induced turbulence | Uniform | [0.25, 0.75] | 0.45 |
| $C_{wl}$ | Wall lubrication force | Uniform | [-0.03, 0] | -0.025 |

As the epistemic uncertainty are quantified and reduced as we introduce new "knowledge" in the form of experimental measurements, the aleatoric uncertainty introduced by the stochastic fluctuation of the two-phase flow system is irreducible. In this work, we evaluate the aleatoric uncertainties in the bubbly flow systems by measuring the fluctuation of mass flow rate, gas injection rate, and bubble sizes. These aleatoric uncertainties influence inlet conditions including phasic velocities, void fraction, and bubble sizes. In practice, we assume the aleatoric uncertainty follow normal distributions with zero mean and a given standard deviation from experimental measurements. In this work, we do not consider the interfacial area transport, instead we assume a constant bubble size as prescribed in the inlet condition. Such an assumption is considered reasonable as we observe only small bubble size change in the measurement.

**4.3 Overall workflow**

Following the proposed modular Bayesian approach, the inverse UQ is implemented following the workflow summarized in Figure 8. STAR-CCM+ is first used to run 500 simulations for surrogate model construction. In these simulations, the constitutive relation parameters are randomly sampled following their prior distribution. The results at the Port 3 plane of these simulations are extracted, reshaped, and combined to form a matrix. PCA is performed on the matrix to identify the first few PCs that can explain more than 90% of the total variance. As illustrated in Figure 2, FNN is then used to take these constitutive relation parameters as inputs to predict the reduced PC subspace outputs, which are then converted back to the original physical space. The combination of PCA and FNN serves as a surrogate model that minimizes both the error of FNN and the error introduced by the PCA.

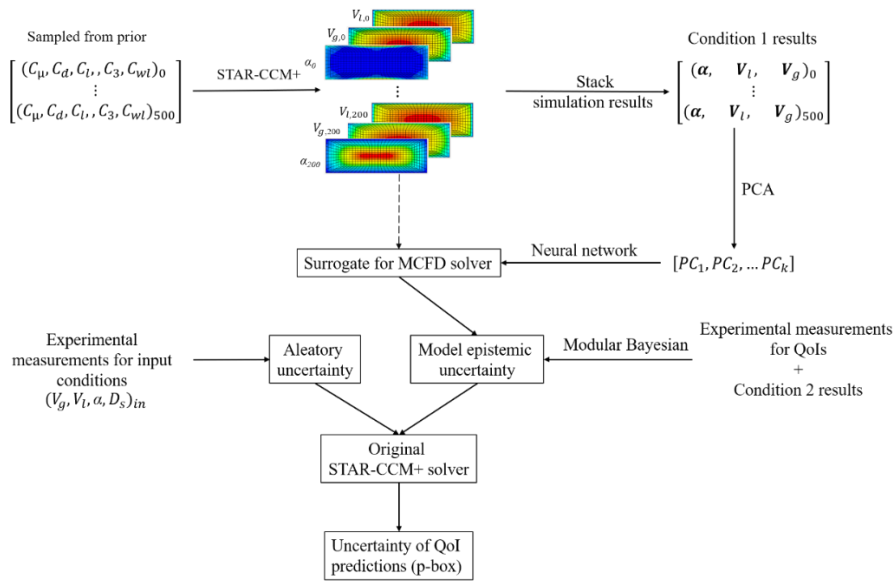

**Figure 8. Workflow of the inverse UQ process**

Based on the surrogate model, modular Bayesian inference is performed for the inverse UQ. Specifically, we use measurements at Run 1 for model parameter uncertainty quantification and use measurements at Run 2 for model form uncertainty quantification. The quantified epistemic uncertainties are then combined with the aleatoric uncertainty of inlet flow conditions to generate samples of both parameters and inlet conditions. These samples are propagated through STAR-CCM+ to generate an ensemble of simulation results. These simulation results considered all the sources of uncertainty are then used to construct p-boxes.

In the UQ process, all three QoIs are normalized to their averaged inlet values to ensure equal importance of the QoIs in the Bayesian inference process as well as the data consistency in the two experiments.

## 5. Results and discussion

### 5.1 Surrogate modeling

As mentioned in Section 2.1, the surrogate model is based on 500 MCFD simulations on Condition 1 with perturbed constitutive relation parameters generated from their prior distributions. The ensemble of these simulations results at $y = 0$, 2, and 4 mm, in comparison to experimental measurements, are depicted in Figure 9. It can be found that the void fraction simulation results are very sensitive to the constitutive relation parameters. In some cases, the void fraction concentrated in the near wall region; while in some other cases, the void fraction concentrated in the central region. Such observations are consistent with the role of interfacial forces. The lift force and wall lubrication force are acted on the lateral direction of the bubbly flow: wall lubrication force pushes the bubble away from the wall, while lift force could either push the bubble away from the wall or push it towards the wall. If the force on one direction is too strong, the bubble would be forced to concentrate on either near wall region or in the central region of the channel. One of the purposes of the UQ process is to identify a reasonable distribution of these parameters so that the lift force and wall lubrication force can act in a balanced way that results in a smooth distribution of void fraction in the lateral direction.

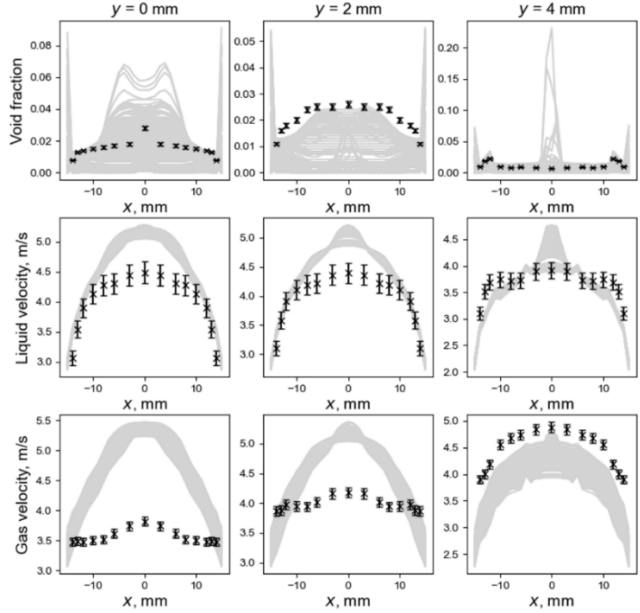

**Figure 9. Ensemble of 500 MCFD simulation results based on parameters sampled from prior distributions, compared with experimental measurements (Condition 1)**

The QoIs at the measurement plane are extracted from each simulation. The results are then reshaped and concatenated to form a matrix in the form of PyTorch tensor, based on which the PCA is performed. In this work, a total of 37 PCs are able to explain more than 90% of the total variance of the full simulation results, thus significantly reduced the dimensionality of the problem. The FNN takes the five empirical parameters as inputs to predict the 37 PCs, the results then are converted back to its original physical space for training. All the training is performed on GPU to ensure computational efficiency.

To mitigate the overfitting issue of FNN, the full dataset is decomposed into two parts: a training dataset consists of 350 randomly chosen results, a testing dataset with the rest 150 results. Training of the FNN is based on the training dataset, while we use only the testing dataset to evaluate the performance of the FNN model. Furthermore, an additional regularization term is introduced in the loss function to further minimize the overfitting issue:

$$L(\hat{y}, y) = \sqrt{\|\hat{y} - y\|_2} + \lambda w^T w, \tag{15}$$

where $w$ is the learnable weights of the FNN. Multiple training cases are performed with different FNN hyperparameter setup, including the number of hidden layers, the number of hidden neuros of each layer, learning rate, and regularization factor $\lambda$, etc. A baseline model is chosen with minimum RMSE and maximum R-square on the testing dataset. The latter describes the ratio between the variance explained by the surrogate model to the total variance of the dataset, and can be computed as:

$$R^2 = 1 - \frac{\sum_{i=1}^{m}(\hat{y}_i - y_i)^2}{\sum_{i=1}^{m}(\bar{y}_i - y_i)^2},\tag{16}$$

where $\bar{y}_i$ is the mean value of the QoIs in the dataset (noted that the three QoIs are all normalized). $R^2$ close to 1 indicates good regression of the model. In this work, the FNN baseline model reaches low RMSE (0.0325) and high $R^2$ (0.996) after training on a few hundred epochs, as depicted in Figure 10.

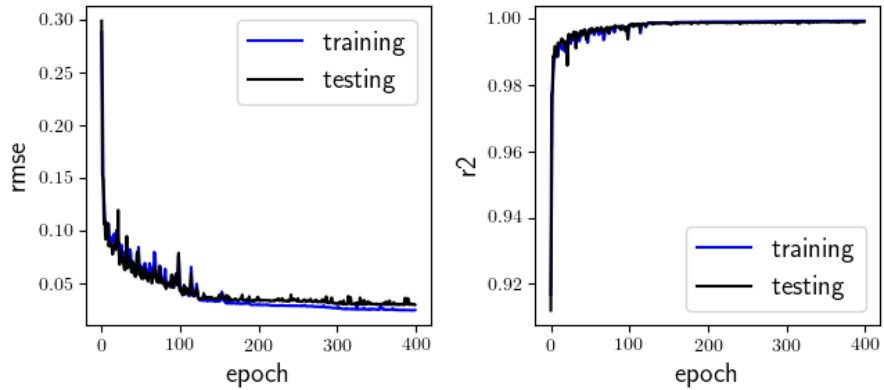

**Figure 10. Root-of-mean-square error and R-square**

The performance of the FNN surrogate model can be further validated against the original MCFD simulation results in the testing dataset, as depicted in Figure 11. It can be found that the FNN predictions on liquid and gas velocity have very good agreement with the original MCFD results. While there are a few instances in void fraction predictions that showing relatively large discrepancies. Such a discrepancy is mainly due to the sensitivity of void fraction prediction to the input interfacial force parameters: a small variation in input parameters may result in a large variation in void fraction prediction. In general, the overall predictions of void fraction can still be regarded as in reasonably good agreement with the MCFD predictions. In this sense, we consider the developed baseline model can be used as a qualified surrogate model in the following UQ process.

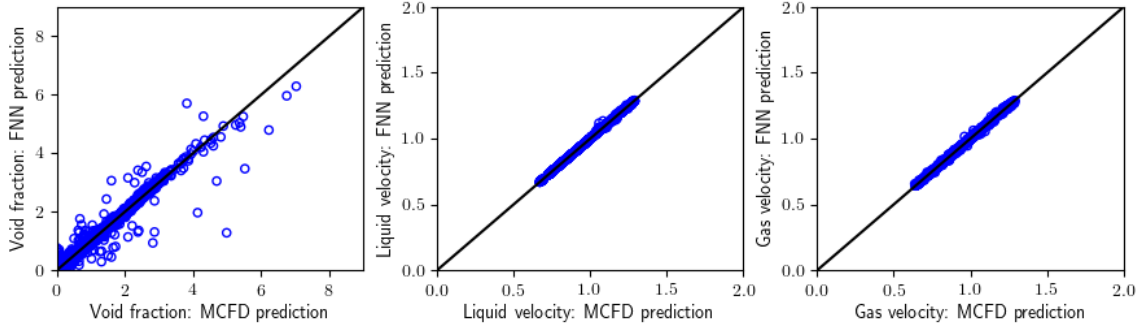

**Figure 11. Validation of neural network predictions against original MCFD simulation results in testing dataset**

### 5.2 Model form uncertainty

As discussed in Section 2.2, the model form uncertainty $\delta(x)$ is modeled with GPs $\mathcal{GP}(0, C)$. It takes the input of local coordinate (*x, y*) to predict the model discrepancy at that location. The GPs are trained with the results from Condition 2. A baseline MCFD simulation with nominal parameter values is performed on Condition 2, and the discrepancy between the simulation results and experimental measurements are obtained at the specific measurement sites to train the GPs. Noted that the training is performed on normalized data, and by doing so, we assume that the obtained $\delta(x)$ can serve as a universal approximator of model form uncertainty and can be applied to other bubbly flow cases, such as Condition 1 case.

In this work, we use Pyro [50], a universal probabilistic programming library based on PyTroch to train the GPs. The uncertainties of the experimental measurements are considered as noise in the training process. We rely on Pyro's built-in optimization algorithm to obtain the maximum a posteriori (MAP) estimate of variance $\sigma^2$ of the covariance kernel of Eq.(7), while manually tuned the length scale $\ell$ to eliminate the unphysical oscillation of the regression results. Three different GPs are trained, i.e. $\delta_\alpha$, $\delta_{Vl}$, and $\delta_{Vg}$, each represents one QoI of the MCFD simulation. The mean and 95% CI of the trained GPs for Condition 2 on *y*= 0, 2, and 4 mm, in comparison to the original model discrepancy, are depicted in Figure 12.

It can be found that both the means and 95% CIs of the three $\delta(x)$ demonstrate smooth pattern without any unphysical oscillations. This is achieved by manually tuning the length scale $\ell$ to a level that two inputs remain correlated until they are very far from each other. It is also observed that the data uncertainty has significant influence on $\delta(x)$. The 95% CIs of the trained $\delta(x)$ are consistent with the measurement uncertainty: larger uncertainty leads to broader CI. For $\delta_{Vl}(x)$ at *y*=4 mm, the near wall region, we observe an almost constant GP mean prediction. This is mainly due to the overall good agreement between MCFD predictions and experimental measurements, as well as the relatively large measurement uncertainty at this location. As shown in Eq.(9), the obtained $\delta(x)$ will serve as a component of the likelihood function in the following Bayesian inference.

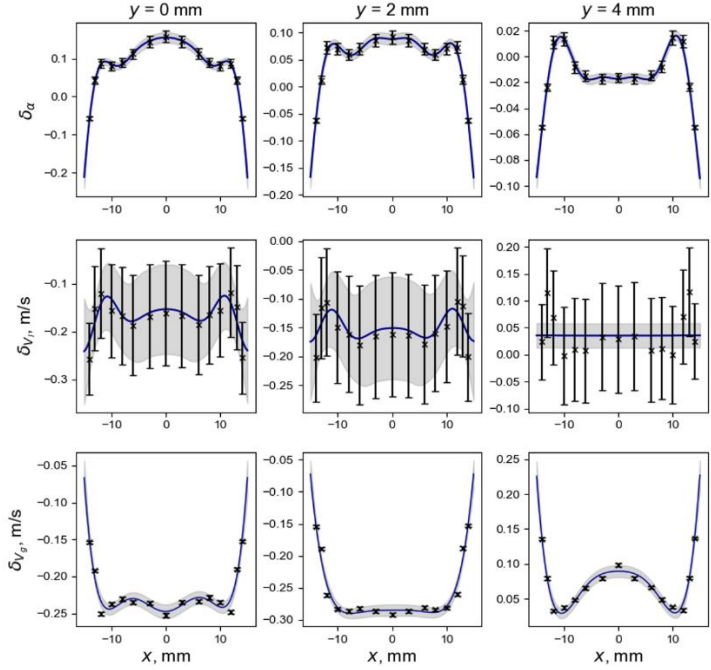

**Figure 12. Mean (in blue line) and 95% CI (in grey area) of the model $\delta(x)$, in comparison to the measured model discrepancy of Condition 2**

### 5.3 Model parameter uncertainty

As discussed in Section 2.2, we rely on MCMC to obtain a stationary Markov chain whose sample can be used to construct the parameter posterior distribution. Such a process requires drawing tens of thousands of samples. In this work, we use the No-U-Turn Sampler (NUTS) algorithm [51] built-in in Pyro to construct the chain. The whole process is performed based on the PyTorch tensor and is computed on GPU to ensure computational efficiency.

In practice, the first 1000 samples from the chain served as "warm-up" samples and are disregarded, after that the obtained samples are regarded as stationary. The "thinning" trick is also applied to the chain that only every 10$^{th}$ element of a chain is used to construct the posterior distribution. Such operation ensures the obtained chain satisfy the definition of Markov chain: sample $t+1$ only dependent on the sample $t$. Such a criterion is also examined by calculating the autocorrelation of the samples. Using the NUTS algorithm, 11,000 samples are drawn, which lead to 1000 "effective" samples to construct the posterior distribution. The traces and autocorrelations of these "effective" samples are depicted in Figure 13. Good mixing and the fast decay of auto-correlations for all parameters are observed. This indicates that the constructed chains can be regarded as stationary

Markov chains, and the samples in the chain can be used to construct the posterior distributions of parameters.

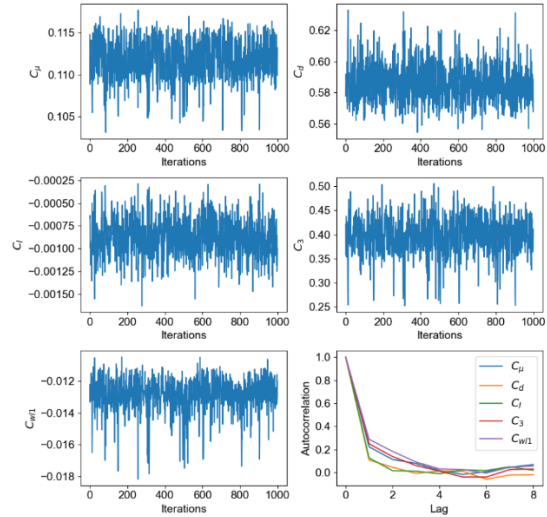

**Figure 13. MCMC sample traces and chain autocorrelations of constitutive relation parameters**

The obtained marginal and point-wise distributions of the three parameters are depicted in Figure 14. The figure shows that the obtained posterior uncertainties of all parameters generally demonstrate a normal distribution pattern, and have a much narrow range compared to their uniform prior distributions summarized in Table 3. It can also be found in the figure that there exist correlations between certain parameter pairs: positive correlation between $C_\mu$ and $C_{wl,1}$, $C_\mu$ and $C_3$, $C_3$ and $C_{wl,1}$; and negative correlation between $C_d$ and $C_{wl,1}$, $C_d$ and $C_{w3}$, $C_d$ and $C_\mu$. Such correlations imply there exist real interactions between the physical phenomena that these constitutive relations represent. The statistics of these parameters, including mean, standard deviation, and 95% HPD, are summarized in Table 4.

**Table 4. posterior statistics of selected empirical parameters**

|  | Mean | Standard deviation | 95% HPD |
|---|---|---|---|
| $C_\mu$ | 0.1116 | 0.0024 | [0.1074, 1.1644] |
| $C_d$ | 0.5862 | 0.0125 | [0.5619, 0.6093] |
| $C_l$ | -0.0008 | 0.0002 | [-0.0013, -0.0004] |
| $C_3$ | 0.3960 | 0.0399 | [0.3251, 0.4755] |
| $C_{wl}$ | -0.0129 | 0.0010 | [-0.0149, -0.0111] |

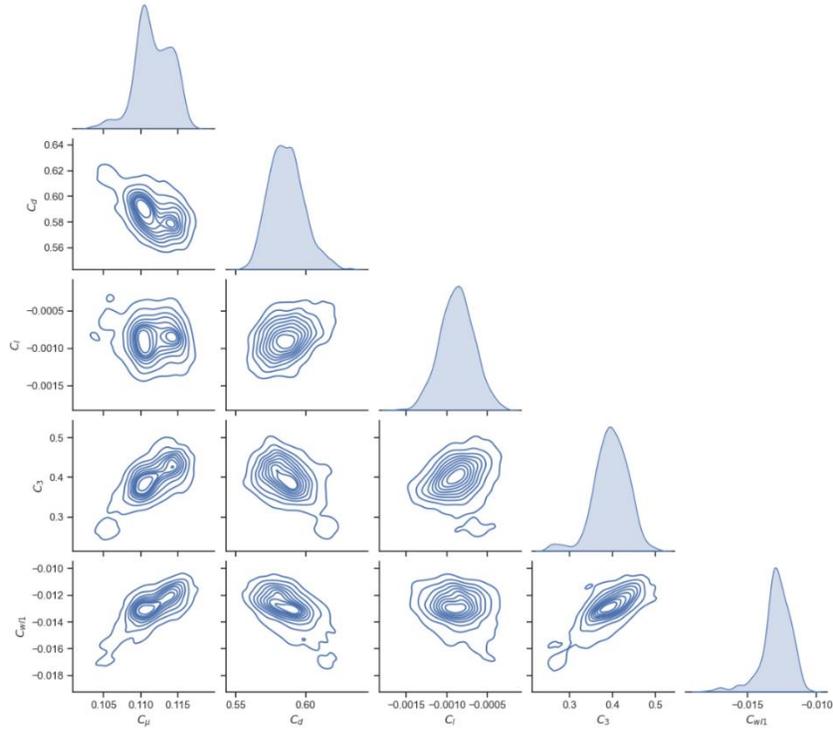

**Figure 14. Marginal and pairwise joint distributions of constitutive relation parameters**

### 5.4 Uncertainty evaluation of MCFD simulation

After the parameter uncertainty and model form uncertainty obtained, a comprehensive uncertainty evaluation of the MCFD simulation can be performed. For the evaluation of epistemic uncertainty only, we run MCFD predictions using STAR-CCM+ on Condition 1 with the constitutive relation parameters drawn from their posterior distributions. The model form uncertainty terms $\delta(x)$ obtained from Condition 2 are then evaluated at every grid point and are added to the simulation results.

To demonstrate the influence of model form uncertainty, QoI predictions with and without the consideration of $\delta(x)$ are depicted respectively, as shown in Figure 15 and Figure 16. It can be found that compared to simulations with prior samples (as shown in Figure 9), the simulations with posterior samples have a much narrower uncertainty distribution. Especially for the void fraction prediction, where the unphysical concentration of void fraction in the near wall region and central region found in the prior simulations both disappeared. This suggests that even without considering model form uncertainty, quantifying the model parameter uncertainty alone could help reduce the uncertainty of MCFD predictions. On the other hand, it should be noted that although the simulation could be improved with posterior parameters, relatively large discrepancies still exist in the prediction. The MCFD prediction underestimates the overall void fraction, especially the central peak (at $x = 0$ and $y = 0$), and corner peak (near the wall region at $y = 4$ mm). Correspondingly, it overestimates the gas phase velocities except for the near wall region at $y = 4$ mm.

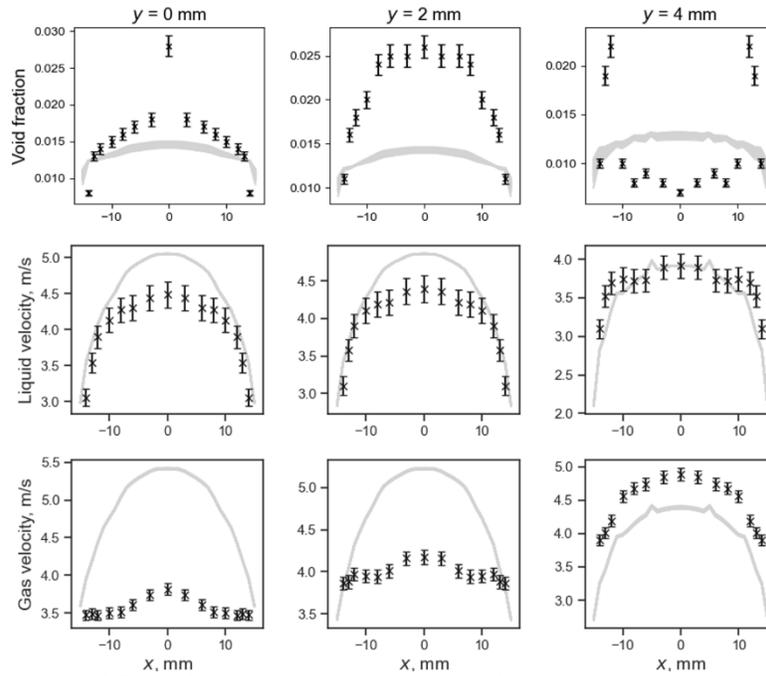

**Figure 15. Ensemble of 50 MCFD simulation results with parameters drawn from posterior distributions, compared against experimental measurements (Condition 1)**

As found in Figure 16, the MCFD predictions can be further improved with the consideration of model form uncertainty. The void fraction prediction showed a better agreement with the measurements, the central peak can be correctly captured. The discrepancies in liquid velocity and gas velocity are also reduced. It also should be noted that while demonstrating better agreement with experimental measurements, the liquid phase prediction actually shows a broader uncertainty range compared to the situation that considers model parameter uncertainty only. This is due to the relatively large uncertainty in the liquid velocity measurements that lead to a $\delta_{Vl}(x)$ with larger uncertainty.

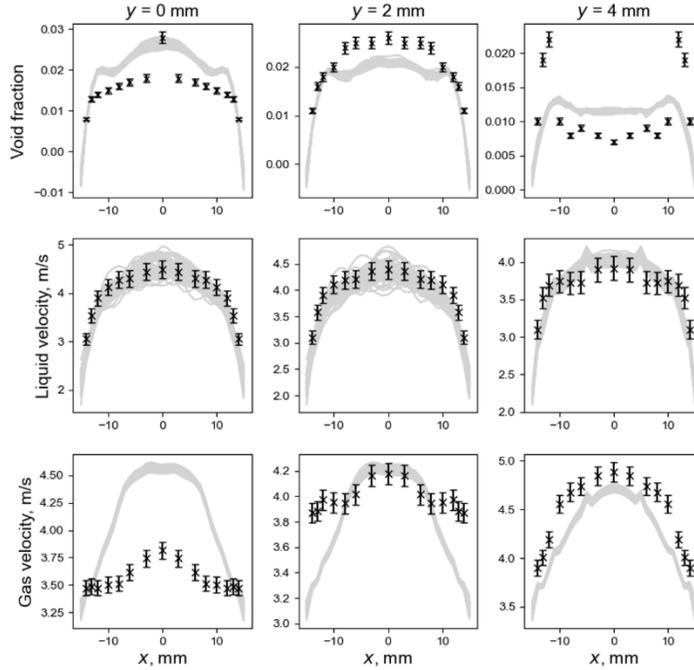

**Figure 16. Ensemble of 50 MCFD posterior simulation combined with the model form uncertainty term, compared against experimental measurements (Condition 1)**

The results clearly demonstrate that with the comprehensive UQ performed based on the proposed approach, the uncertainty of MCFD predictions can be significantly reduced. However, it should also be noted that even after the UQ, discrepancies still exist between MCFD predictions and experimental measurements. The most significant discrepancies can be found in void fraction prediction at near wall region ($y = 4$ mm), and gas velocity prediction at $y = 0$ and 2 mm regions. We believe the most important reason that caused such discrepancies comes from the model form uncertainty term $\delta(x)$ modeled by GPs. In this work, we assume $\delta(x)$ is solely dependent on locations but independent of inlet flow conditions $\eta$. Under this assumption, we train $\delta(x)$ on normalized data and make it a universal approximator for any bubbly flows in the rectangular channel. However, in reality $\delta(x)$ should also be dependent on inlet conditions, including void fraction, phasic velocities, and bubble size. In this sense, $\delta(x)$ should be expanded to $\delta(x, \eta)$. This approach, however, would require a much larger experiment database with measurements performed on multiple inlet flow conditions. Further investigation on a more accurate model form uncertainty model would be required to further improve the work.

The final step for the comprehensive UQ process is to combine epistemic uncertainty with aleatoric uncertainty to construct p-boxes. As discussed in Section 4.2, the aleatoric uncertainty would cause fluctuating inlet flow conditions, including two phasic velocities, void fraction, and bubble sizes. In this work, we assume the aleatoric uncertainty follow normal distributions with zero mean and standard deviations that can be derived from Table 1. We sample 20 samples with perturbing inlet flow conditions from the aleatoric uncertainty distribution. For each of these samples with a prescribed inlet flow condition, we draw 10 samples from parameter posterior distribution and model form uncertainty distribution. A total 200 simulation is performed for p-boxes construction. An example constructed at the center of the channel is depicted in Figure 17, along with the comparison to the 95% CI of the measurements. The constructed p-boxes can serve as validation metrics to quantitatively evaluate the agreement between model prediction and experimental measurements. Furthermore, it can be used for a more comprehensive analysis within the framework of uncertainty and risk, including risk assessment [39], reliability analysis [40], and sensitivity analysis [41].

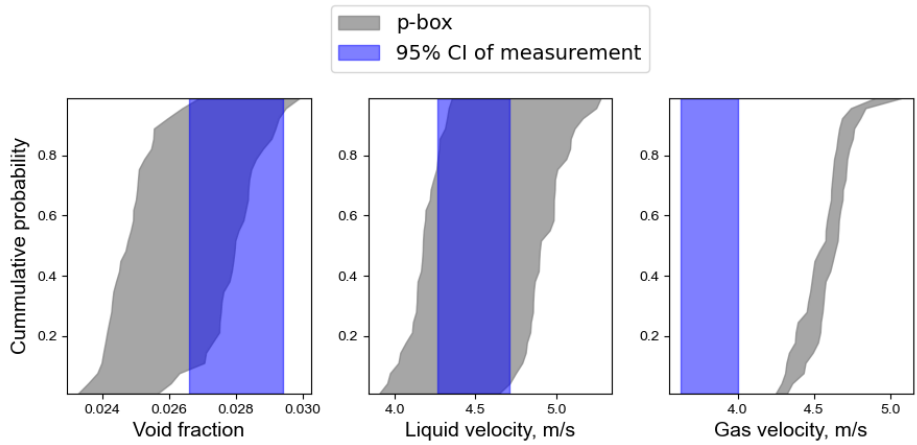

**Figure 17. p-boxes of three QoIs at $x = 0$ mm, $y = 0$ mm, in comparison to experimental measurement (Condition 1).**

## 6. Conclusions

In this work, we introduce a modular Bayesian approach to quantify and reduce the uncertainties of Multiphase Computational Fluid Dynamics (MCFD) simulations. The proposed approach is supported by three machine learning methods: principal component analysis for dimensionality reduction, feedforward neural network for surrogate modeling, and Gaussian processes for model form uncertainty evaluation. The epistemic uncertainty is quantified through the modular Bayesian approach, then combined with the aleatoric uncertainty of the stochastic flow fluctuation for a comprehensive uncertainty evaluation of the MCFD prediction. Based on the obtained uncertainty, probability-boxes can be constructed for comprehensive risk analysis.

With the support of high-resolution experimental measurements at a 10mm×30mm square channel, a comprehensive case study has been performed to evaluate the uncertainty of three quantities of interest: void fraction, liquid velocity, and gas velocity. The proposed approach is implemented based on the open-source deep learning library PyTorch and can be run on GPU for efficient and fast evaluation. The case study demonstrates that the proposed approach can effectively quantify and reduce the uncertainty of MCFD predictions. With the consideration of model parameter uncertainty only, the prediction uncertainty can be significantly reduced, and the unphysical predictions can be eliminated. The results can be further improved with the combination of model parameter uncertainty and model form uncertainty. The proposed approach provides a general framework that utilizing different experimental measurements for a comprehensive UQ of high-resolution, low-fidelity numerical models.

Be that as it may, it should also be admitted that discrepancies between MCFD predictions and experimental measurements still exist after the UQ is performed. The main reason is the model form uncertainty form trained with Gaussian process neglects the influence of inlet conditions. To further improve the work, additional experiments need to be performed on a broader range of inlet conditions. Based on the larger measurement database, a model form uncertainty that takes both location and inlet conditions as input can be trained to better reflect the physics underlying the bubbly flow system.

**Acknowledgement**

This material is based upon work supported by Laboratory Directed Research and Development (LDRD) funding from Argonne National Laboratory, provided by the Director, Office of Science, of the U.S. Department of Energy under Contract No. DEAC02-06CH11357.

The initial version of this work is supported by the U.S. Department of Energy Office of Nuclear Energy's Nuclear Energy University Program via the Integrated Research Project on "Development and Application of a Data-Driven Methodology for Validation of Risk-Informed Safety Margin Characterization Models" under Contract No. DE-NE0008530.

The experiment in this work is performed using funding received from the DOE Office of Nuclear Energy's Nuclear Energy University Program under Contract No. DE-NE0008535.

**References**

1. Zugazagoitia E., Queral C., Fernández-Cosials K., Gómez J., Durán L.F., Sánchez-Torrijos J., Posada J.M., 2020. Uncertainty and sensitivity analysis of a PWR LOCA sequence using parametric and non-parametric methods. Reliab. Eng. Syst. Saf. 193, 106607.

2. Kang D.G., 2020. Comparison of statistical methods and deterministic sensitivity studies for investigation on the influence of uncertainty parameters: Application to LBLOCA. Reliab. Eng. Syst. Saf. 203, 107082.

3. Galushin S., Grishchenko D., Kudinov P., 2020. Implementation of Framework for Assessment of Severe Accident Management Effectiveness in Nordic BWR. Reliab. Eng. Syst. Saf. , 107049.

4. Bodda S.S., Gupta A., Dinh N., 2020. Enhancement of risk informed validation framework for external hazard scenario. Reliab. Eng. Syst. Saf. 204, 107140.

5. Saini N. and Bolotnov I.A., 2020. Interface capturing simulations of droplet interaction with spacer grids under DFFB conditions. Nucl. Eng. Des. 364, 110685.

6. Colombo M. and Fairweather M., 2016. Accuracy of Eulerian–Eulerian, two-fluid CFD boiling models of subcooled boiling flows. Int. J. Heat Mass Transfer. 103, 28-44.

7. Krepper E., Končar B., Egorov Y., 2007. CFD modelling of subcooled boiling—concept, validation and application to fuel assembly design. Nucl. Eng. Des. 237(7), 716-731.

8. Ishii M. and Hibiki T., 2011. Thermo-Fluid Dynamics of Two-Phase Flow. Springer.

9. Liu Y. and Dinh N., 2018. Flow Boiling in Tubes. Book chapter in: Handbook of Thermal Science and Engineering. Springer, Cham.

10. Fu Y. and Liu Y., 2016. Development of a robust image processing technique for bubbly flow measurement in a narrow rectangular channel. Int. J. Multiphase Flow. 84, 217-228.

11. Zhou X., Doup B., Sun X., 2013. Measurements of liquid-phase turbulence in gas–liquid two-phase flows using particle image velocimetry. Measurement Science and Technology. 24(12), 125303.

12. Shi S., Wang D., Qian Y., Sun X., Liu Y., Tentner A., 2020. Liquid-phase turbulence measurements in air-water two-phase flows using particle image velocimetry. Prog. Nuclear Energy. 124, 103334.

13. Wang D., Song K., Fu Y., Liu Y., 2018. Integration of conductivity probe with optical and x-ray imaging systems for local air–water two-phase flow measurement. Measurement Science and Technology. 29(10), 105301.

14. He X., Zhao F., Vahdati M., 2020. Uncertainty Quantification of Spalart–Allmaras Turbulence Model Coefficients for Simplified Compressor Flow Features. Journal of Fluids Engineering. 142(9)

15. Wu X., Liu Y., Kearfott K., Sun X., 2020. Evaluation of public dose from FHR tritium release with consideration of meteorological uncertainties. Sci. Total Environ. 709, 136085.

16. Gong H., Yu Y., Li Q., Quan C., 2020. An inverse-distance-based fitting term for 3D-Var data assimilation in nuclear core simulation. Ann. Nucl. Energy. 141, 107346.

17. Wu X., Kozlowski T., Meidani H., Shirvan K., 2018. Inverse uncertainty quantification using the modular Bayesian approach based on Gaussian process, Part 1: Theory. Nucl. Eng. Des. 335, 339-355.

18. Liu Y. and Dinh N., 2019. Validation and Uncertainty Quantification for Wall Boiling Closure Relations in Multiphase-CFD Solver. Nucl. Sci. Eng. 193(1-2), 81-99.

19. Nguyen H.D. and Gouno E., 2020. Bayesian inference for Common cause failure rate based on causal inference with missing data. Reliab. Eng. Syst. Saf. 197, 106789.

20. Kennedy M.C. and O'Hagan A., 2001. Bayesian calibration of computer models. J. Royal Stat. Soc: Series B (Statistical Methodology). 63(3), 425-464.

21. Wang C., Wu X., Kozlowski T., 2017. Surrogate-Based Inverse Uncertainty Quantification of TRACE Physical Model Parameters Using Steady-State PSBT Void Fraction Data. Proc.17th Int.Topl.Mtg.Nuclear Reactor Thermal Hydraulics (NURETH-17). , 3-8.

22. Wang C., Wu X., Kozlowski T., 2019. Gaussian Process–Based Inverse Uncertainty Quantification for TRACE Physical Model Parameters Using Steady-State PSBT Benchmark. Nucl. Sci. Eng. 193(1-2), 100-114.

23. Radaideh M.I., Borowiec K., Kozlowski T., 2019. Integrated framework for model assessment and advanced uncertainty quantification of nuclear computer codes under Bayesian statistics. Reliab. Eng. Syst. Saf. 189, 357-377.

24. Liu Y., Sun X., Dinh N.T., 2019. Validation and uncertainty quantification of multiphase-CFD solvers: A data-driven Bayesian framework supported by high-resolution experiments. Nucl. Eng. Des. 354, 110200.

25. Sun D., Wainwright H.M., Oroza C.A., Seki A., Mikami S., Takemiya H., Saito K., 2020. Optimizing long-term monitoring of radiation air-dose rates after the Fukushima Daiichi Nuclear Power Plant. J. Environ. Radioact. 220-221, 106281.

26. Radaideh M.I. and Kozlowski T., 2020. Surrogate modeling of advanced computer simulations using deep Gaussian processes. Reliab. Eng. Syst. Saf. 195, 106731.

27. Li M., Sadoughi M., Hu Z., Hu C., 2020. A hybrid Gaussian process model for system reliability analysis. Reliab. Eng. Syst. Saf. 197, 106816.

28. Wu X., Kozlowski T., Meidani H., 2018. Kriging-based inverse uncertainty quantification of nuclear fuel performance code BISON fission gas release model using time series measurement data. Reliab. Eng. Syst. Saf. 169, 422-436.

29. El Moçayd N., Shadi Mohamed M., Ouazar D., Seaid M., 2020. Stochastic model reduction for polynomial chaos expansion of acoustic waves using proper orthogonal decomposition. Reliab. Eng. Syst. Saf. 195, 106733.

30. Paszke A., Gross S., Chintala S., Chanan G., 2017. PyTorch: Tensors and dynamic neural networks in Python with strong GPU acceleration.

31. Rumelhart D.E., Hinton G.E., Williams R.J., 1986. Learning representations by back-propagating errors. Nature. 323(6088), 533-536.

32. Bao H., Feng J., Dinh N., Zhang H., 2020. Computationally Efficient CFD Prediction of Bubbly Flow using Physics-Guided Deep Learning. Int. J. Multiphase Flow. , 103378.

33. Liu Y., Dinh N., Sato Y., Niceno B., 2018. Data-driven modeling for boiling heat transfer: using deep neural networks and high-fidelity simulation results. Appl. Therm. Eng. 144, 305-320.

34. Bao H., Dinh N., Lin L., Youngblood R., Lane J., Zhang H., 2020. Using deep learning to explore local physical similarity for global-scale bridging in thermal-hydraulic simulation. Ann. Nucl. Energy. 147, 107684.

35. Rasmussen C.E., 2004. Gaussian processes in machine learning. Book chapter in: Advanced lectures on machine learning. Springer.

36. Liu Y., Dinh N.T., Smith R.C., Sun X., 2019. Uncertainty quantification of two-phase flow and boiling heat transfer simulations through a data-driven modular Bayesian approach. Int. J. Heat Mass Transfer. 138, 1096-1116.

37. Higdon D., Nakhleh C., Gattiker J., Williams B., 2008. A Bayesian calibration approach to the thermal problem. Comput. Methods Appl. Mech. Eng. 197(29), 2431-2441.

38. Ferson S. and Oberkampf W.L., 2009. Validation of imprecise probability models. International Journal of Reliability and Safety. 3(1-3), 3-22.

39. Shortridge J., Aven T., Guikema S., 2017. Risk assessment under deep uncertainty: A methodological comparison. Reliab. Eng. Syst. Saf. 159, 12-23.

40. Simon C. and Bicking F., 2017. Hybrid computation of uncertainty in reliability analysis with p-box and evidential networks. Reliab. Eng. Syst. Saf. 167, 629-638.

41. Schöbi R. and Sudret B., 2019. Global sensitivity analysis in the context of imprecise probabilities (p-boxes) using sparse polynomial chaos expansions. Reliab. Eng. Syst. Saf. 187, 129-141.

42. Kim S., Fu X.Y., Wang X., Ishii M., 2000. Development of the miniaturized four-sensor conductivity probe and the signal processing scheme. Int. J. Heat Mass Transfer. 43(22), 4101-4118.

43. Liu Y., Wang C., Qian Y., Sun X., 2020. Uncertainty analysis of PIV measurements in bubbly flows considering sampling and bubble effects with ray optics modeling. Nucl. Eng. Des. 364, 110677.

44. Wang D., Liu Y., Talley J.D., 2018. Numerical evaluation of the uncertainty of double-sensor conductivity probe for bubbly flow measurement. Int. J. Multiphase Flow. 107, 51-66.

45. Bao H., Dinh N.T., Lane J.W., Youngblood R.W., 2019. A data-driven framework for error estimation and mesh-model optimization in system-level thermal-hydraulic simulation. Nucl. Eng. Des. 349, 27-45.

46. Sato Y., Sadatomi M., Sekoguchi K., 1981. Momentum and heat transfer in two-phase bubble flow—II. A comparison between experimental data and theoretical calculations. Int. J. Multiphase Flow. 7(2), 179-190.

47. Antal S.P., Lahey R.T., Flaherty J.E., 1991. Analysis of phase distribution in fully developed laminar bubbly two-phase flow. Int. J. Multiphase Flow. 17(5), 635-652.

48. Launder B.E. and Spalding D.B., 1974. The numerical computation of turbulent flows. Comput. Methods Appl. Mech. Eng. 3(2), 269-289.

49. Troshko A.A. and Hassan Y.A., 2001. A two-equation turbulence model of turbulent bubbly flows. Int. J. Multiphase Flow. 27(11), 1965-2000.

50. Bingham E., Chen J.P., Jankowiak M., Obermeyer F., Pradhan N., Karaletsos T., Singh R., Szerlip P., Horsfall P., Goodman N.D., 2019. Pyro: Deep universal probabilistic programming. The Journal of Machine Learning Research. 20(1), 973-978.

51. Hoffman M.D. and Gelman A., 2014. The No-U-Turn sampler: adaptively setting path lengths in Hamiltonian Monte Carlo. J.Mach.Learn.Res. 15(1), 1593-1623.